\documentclass[twocolumn,aps,showpacs,nofootinbib,prd]{revtex4-1}
\usepackage{amssymb}
\usepackage{amsmath}
\usepackage{graphics}
\usepackage{epsfig}
\usepackage[dvips]{color}

\newcommand{\be}{\begin{equation}}
\newcommand{\ee}{\end{equation}}
\newcommand{\bea}{\begin{eqnarray}}
\newcommand{\eea}{\end{eqnarray}}
\newcommand{\nn}{\nonumber}
\newcommand{\beas}{\begin{eqnarray*}}
\newcommand{\eeas}{\end{eqnarray*}}

\newcommand{\slsh}[1]{{\not \! #1}}

\newcommand{\bd}[1]{{\bf #1}}

\begin{document}

\title{Jet asymmetry and momentum imbalance from $2\rightarrow 2$ and $2\rightarrow 3$ partonic processes \\ in relativistic heavy-ion collisions} 
\author{Alejandro Ayala$^{1,5}$, Isabel Dominguez$^2$, Jamal
  Jalilian-Marian$^3$ and Maria Elena
  Tejeda-Yeomans$^{1,4}$} 
\affiliation{$^1$Instituto de Ciencias
  Nucleares, Universidad Nacional Aut\'onoma de M\'exico, Apartado
  Postal 70-543, M\'exico Distrito Federal 04510,
  Mexico.\\ $^2$Facultad de Ciencias F\'isico-Matem\'aticas, Universidad Aut\'onoma de Sinaloa,
Avenida de las Am\'ericas y Boulevard Universitarios, Ciudad Universitaria,
C.P. 80000, Culiac\'an, Sinaloa, M\'exico.\\ $^3$Department of Natural Sciences, Baruch College, New
  York, New York 10010, USA and CUNY Graduate Center, 365 Fifth
  Avenue, New York, New York 10016, USA.
\\$^4$Departamento de F\'{\i}sica,
  Universidad de Sonora, Boulevard Luis Encinas J. y Rosales, Colonia
  Centro, Hermosillo, Sonora 83000, Mexico.
\\$^5$Centre for Theoretical and Mathematical Physics, and Department of Physics,
  University of Cape Town, Rondebosch 7700, South Africa}

\begin{abstract}

\noindent We study momentum imbalance as a function of jet asymmetry in high-energy heavy-ion collisions. To implement  parton production during the collision, we include all Leading Order (LO) $2\to 2$ and $2\to 3$ parton processes in pQCD. The produced partons lose energy within the quark gluon plasma and hadronize collinearly when they leave it. The energy and momentum deposited into the plasma is described using linear viscous hydrodynamics with a constant energy loss per unit length and a total energy loss given by a Gaussian probability centered around a mean value $\bar{\mathcal{E}}$ and a half-width $\Delta{\mathcal{E}}$. We argue that the shape of the asymmetry observed by the CERN-CMS Collaboration can indeed be attributed to parton energy loss in the medium and that a good description of data is achieved when one includes a slight enhancement coming from the contribution of $2\to 3$ parton processes that modifies the asymmetry distribution of the dijet events. We compare our results to CMS data for the most central collisions and study different values for $\bar{\mathcal{E}}$ and $\Delta{\mathcal{E}}$.
\end{abstract}

\maketitle

\section{Introduction}\label{Introduction}

Strongly interacting hard probes are useful tools to study the properties of the quark-gluon plasma created in a relativistic heavy-ion collision. These hard probes, energetic quarks or gluons, are produced in the hard scattering that occurs at the earliest stage of the collision. They transfer energy to the medium via multiple scatterings and eventually emerge as hadrons. The way the energy is transferred to the medium can be studied to infer some properties of the plasma such as temperature, viscosity, density, size, etc. 
To this end an important observable is the momentum imbalance between leading and away-side jets which allows one to investigate how momentum gets distributed in the away-side. Another such observable is the {\it asymmetry distribution} as a function of the jet asymmetry $A_J$ defined by
\be
\label{asymm}
A_J = \frac{p_{\mathrm T,1} - p_{\mathrm T,2}}{p_{\mathrm T,1}+p_{\mathrm T,2}}
\ee
where $p_{\mathrm T,1}, p_{\mathrm T,2}$ are the transverse momenta of the reconstructed leading jet and subleading jet respectively. The jet asymmetry is therefore positive definite by construction. When both jets have similar momenta, $A_J$ is close to $0$ whereas a monojet (that usually corresponds to the case where the jet in the away-side is totally quenched) has asymmetry $A_J=1$. 

The CMS collaboration has measured this asymmetry for jets with cone radius $R = 0.5$, triggering on leading jets with $p_{\mathrm T,1} > 120$ GeV and on subleading jets with $p_{\mathrm T,1} > 50$ GeV~\cite{CMS1, CMS2}. Taking the case reported in Ref.~\cite{CMS1}, for the most central PbPb events where the data set contains only jets with $p_{\mathrm T,1} \leq 300$ GeV, the asymmetry distribution shows a significant deficit of events with $A_J\sim 0$ as well as a significant excess of events with $A_J\sim 0.4$ as compared to a PYTHIA + HYDJET simulation. This behaviour 
in the measured asymmetry distribution has been studied under different approaches: Monte Carlo simulations with improved jet finding algorithms, parametric studies of energy loss models in connection with jet finding algorithms, models where NLO pQCD behaviour of jet in-medium evolution is incorporated and solving relativistic hydrodynamic equations coupled with hadron momentum distribution profiles.~\cite{renk,majumder,others}. The interplay of jets and the hydrodynamic evolution of the bulk was first discussed in the context of Mach cone formation \cite{jetshydro} and has been extended recently to other observables \cite{jetsother}. The angular broadening of a medium-induced QCD cascade has also been recently studied in Ref.~\cite{Blaizot}

With these studies at hand, recent reviews on the state-of-the-art and future of jet physics in heavy ion collisions, attribute the measured jet asymmetry mostly to energy-loss mechanisms and argue that it is possible to go a step forward and look for other properties that could be constrained by this asymmetry (see for example \cite{renk,majumder} and references therein).

Another observable that can be measured as a function of $A_J$ is the {\it missing transverse momentum}. This is constructed on an event by event basis, projecting the momentum of measured tracks onto the leading jet axis and summing over all tracks with transverse momentum larger than a minimum value $p_{\mathrm T}^{\mathrm min}>0.5$.
\be
\label{ptmiss}
\slsh{p}_{\mathrm T}^{\|}  = \sum_{i} - p_{\mathrm T}^i \cos(\phi_i-\phi_{L}).
\ee
This quantity is negative definite in the leading jet hemisphere and positive definite in the away-side hemisphere~\cite{CMS1, CMS2}. The missing momentum in the away-side can be obtained either for tracks within the reconstructed jet around the subleading hadron, in which case one expects that the momentum outside the jet compensates the {\it missing} one inside the jet, or else, on the whole away-side hemisphere. The data in Ref.~\cite{CMS1} shows that a large negative contribution to the average missing momentum $\langle\slsh{p}_{\mathrm T}^{\|}\rangle$ (in the direction of the leading jet) for $p_{\mathrm T} > 8$ GeV, is balanced mostly by the combined contributions from the 0.5-8 GeV regions outside a cone of angular size $\Delta\phi < \pi /6$ centered along a direction opposite to the leading hadron. There is however also a small contribution outside this cone from tracks with $p_{\mathrm T}>8$ GeV with $0.3<A_J<0.5$.

The question we want to address in this work is whether these two observables, the asymmetry distribution and the missing transverse momentum, can be better characterized if we account for the possible contributions from partonic processes where the hard collision of two partons in the incoming protons/nuclei gives rise to three partons in the final state. Qualitatively, two hard partons in the away-side that subsequently hadronize produce a broader asymmetry distribution and a harder component of the missing momentum outside the reference cone of the subleading jet which goes more in line with the reported measurements. This kind of studies can be important to carry out a more precise determination of the energy loss mechanism.

In two recent works~\cite{Ayala1, Ayala2} we have studied the effect of hadron production from $2\rightarrow 3$ parton processes in azimuthal angular correlations using the leading order QCD matrix elements. These works have been extended to the description of the energy-momentum deposited into the medium~\cite{Ayala3, Ayala4} using linearized viscous hydrodynamics. In this work we apply this formalism to study the relative contribution from $2 \to 3$ and $2 \to 2$ parton initiated jets in the analysis of both the asymmetry distribution and momentum imbalance.  The work is organized as follows: In Sec.~\ref{II} we collect the ingredients needed to describe the energy-momentum deposited in the medium by fast moving partons using linearized viscous hydrodynamics. In Sec.~\ref{III} we present the results of this analysis and show that by including a 20\% contribution from $2\to 3$ parton initiated processes one can achieve a better description of the asymmetry and momentum imbalance distributions. We finally summarize our results and conclude in Sec.~\ref{Conclusions}.

\begin{figure*}
{\centering {
\includegraphics[scale=0.4]{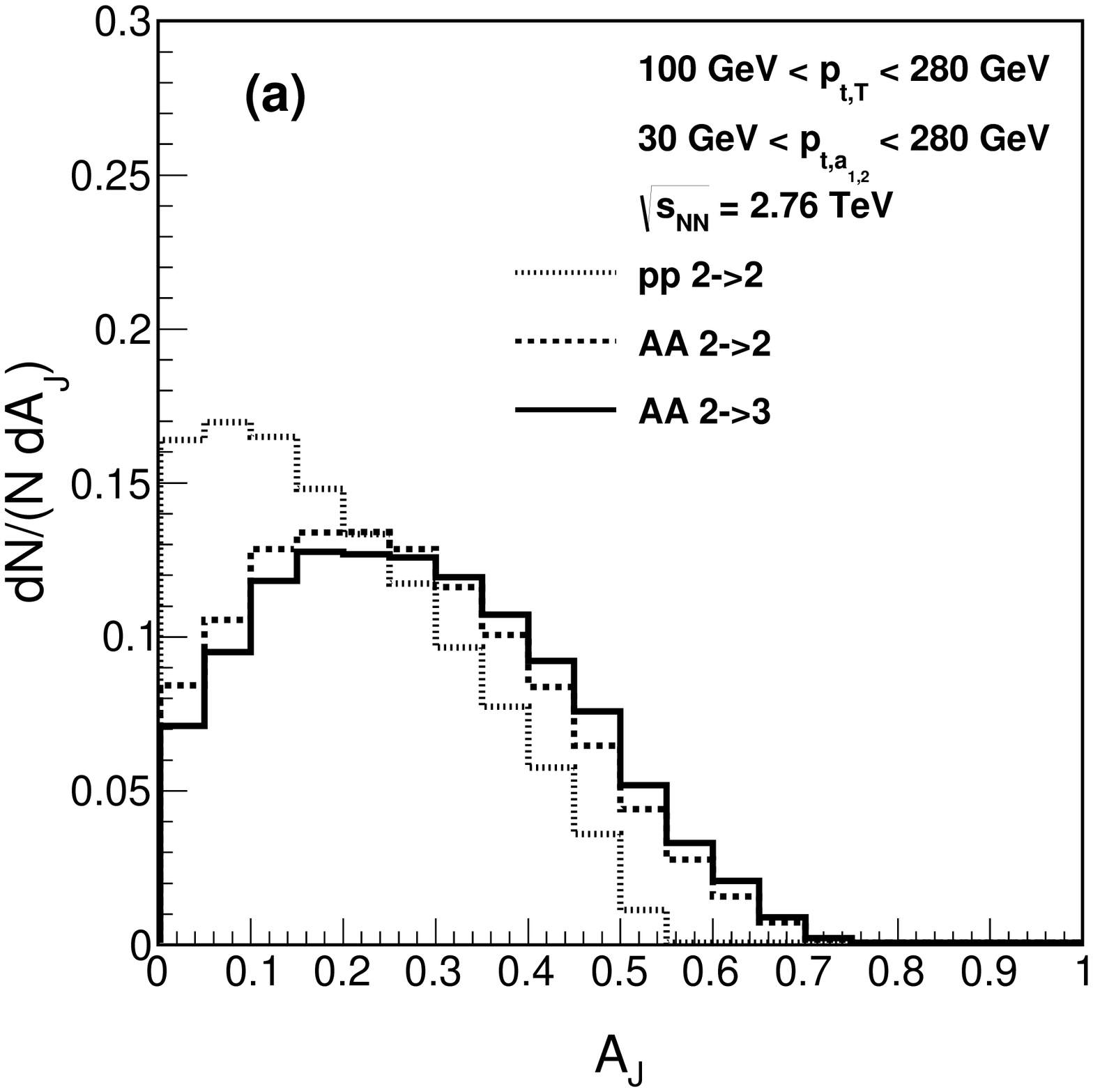}
\includegraphics[scale=0.4]{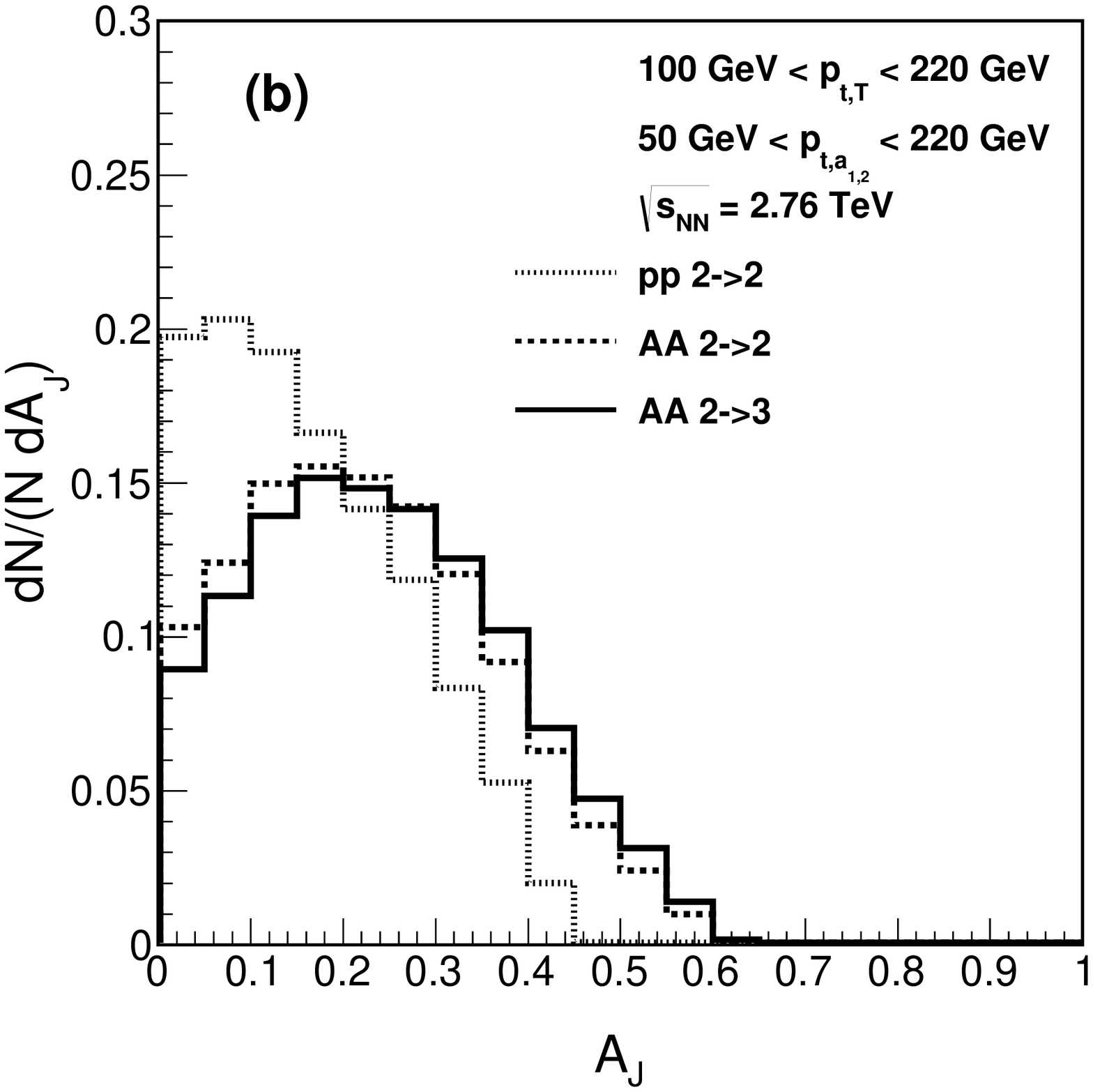}
}}
\caption{Simulated dijet asymmetry distribution for leading jets of $p_{\mathrm T,1} > 100$ GeV and subleading jets of
$p_{\mathrm T,2} > 30$ GeV (left) and $p_{\mathrm T,2} > 50$ GeV (right). The histograms show pp $2\to 2$ (fine dotted line) and AA $2\to 2$ (dotted line) and $2\to 3$ (solid line), separately.}
\label{fig1}
\end{figure*}

\begin{figure*}
{\centering {

\includegraphics[scale=0.4]{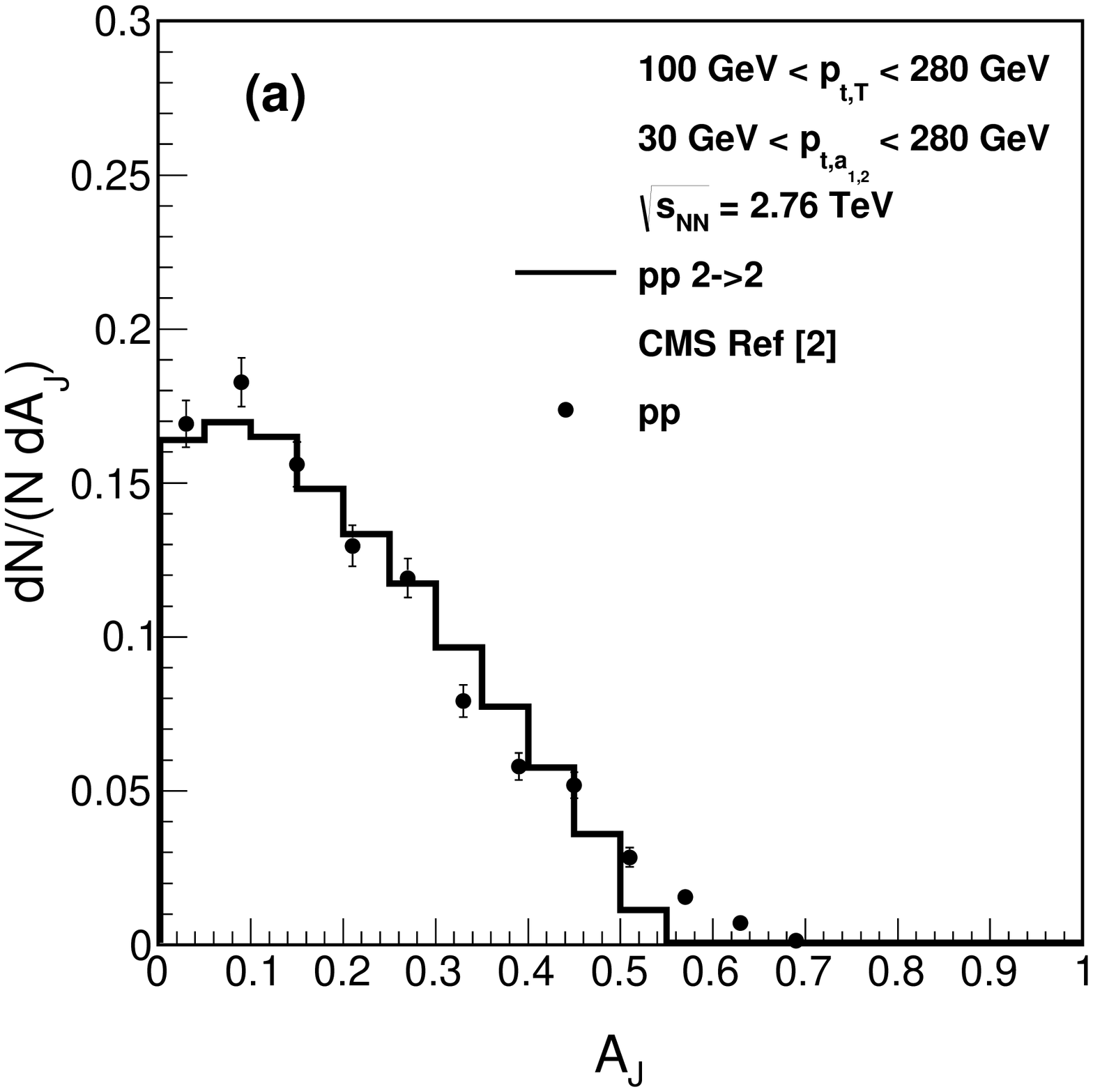}\includegraphics[scale=0.4]{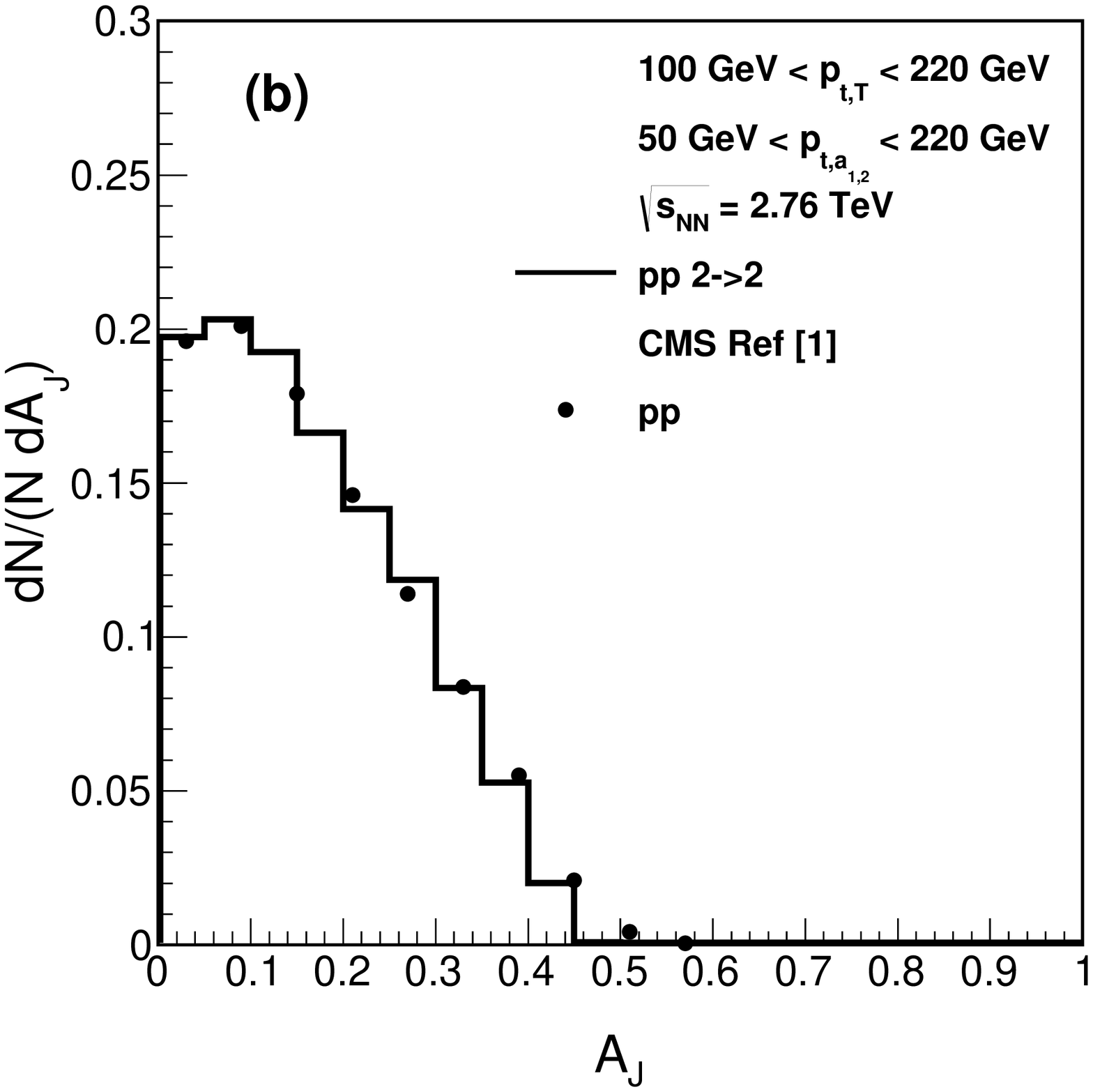}

\includegraphics[scale=0.4]{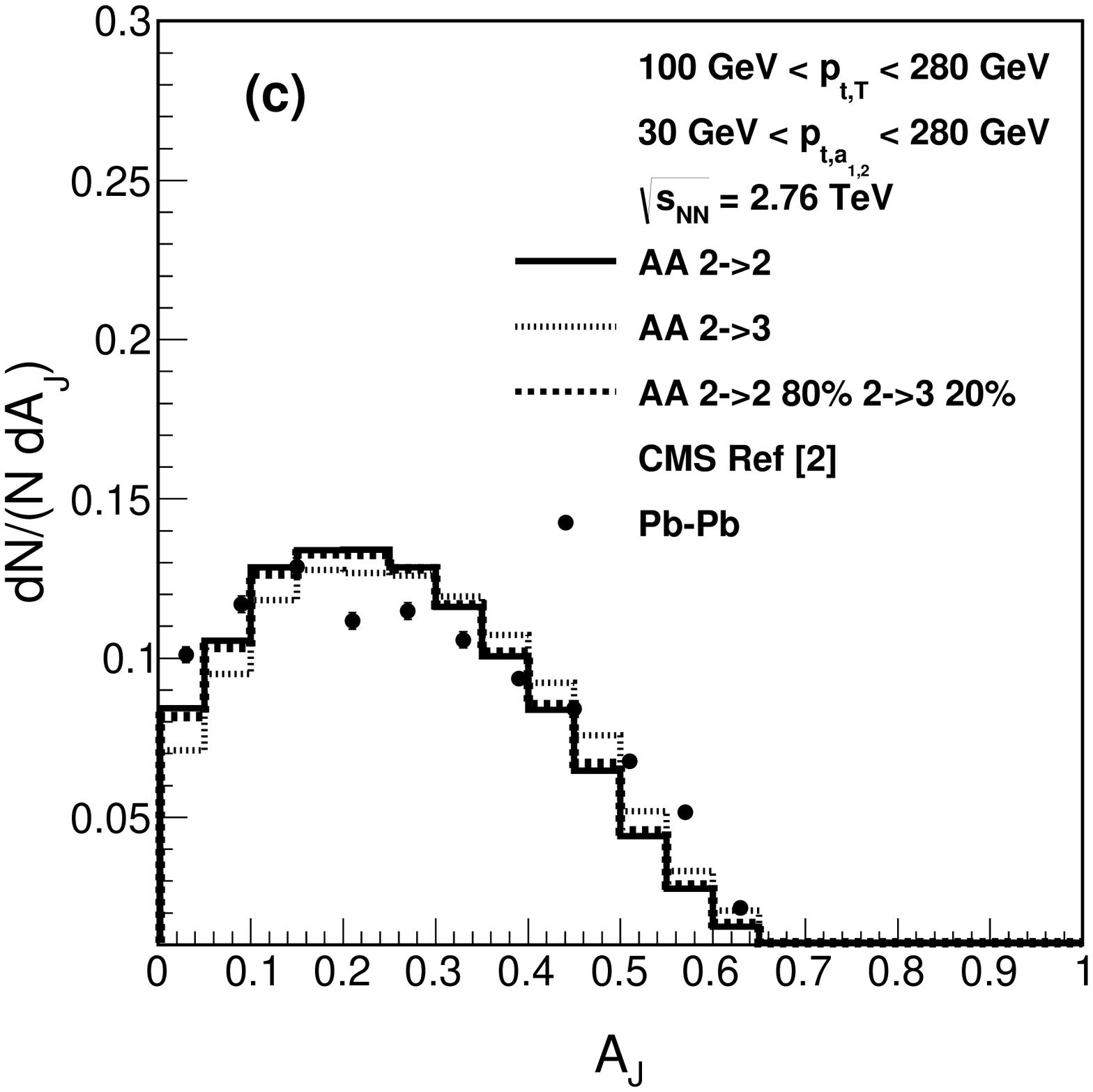}\includegraphics[scale=0.4]{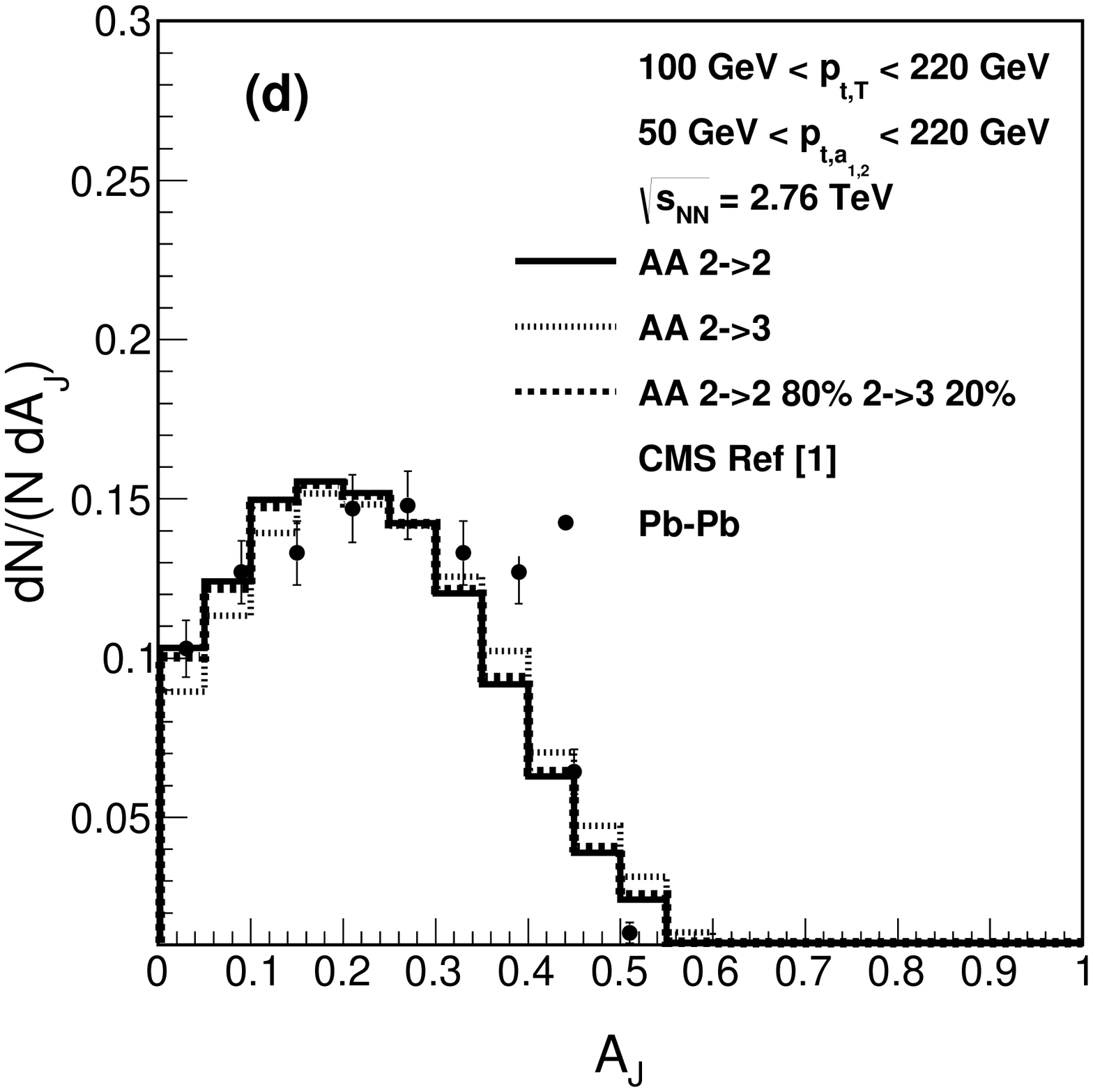}

}}
\caption{
Simulated dijet asymmetry distribution for leading jets of $p_{\mathrm T,1} > 100$ GeV and subleading jets of
$p_{\mathrm T,2} > 30$ GeV (left) and $p_{\mathrm T,2} > 50$ GeV (right)
The histograms in the top row show the benchmark asymmetry distribution for the pp $2\to 2$ sample compared to CMS data for pp collisions in the 0-10\% centrality bin. The histograms on the bottom row show the asymmetry distribution for the AA $2\to 2$ (solid line) $2\to 3$ (fine dotted line) and {\it combined} (dotted line) samples, compared to CMS data for Pb-Pb collisions in the 0-10\% centrality bin. The {\it combined} sample has 80\% of $2 \to 2$ and 20\% of $2 \to 3$ AA events.
}
\label{fig2}
\end{figure*}

\begin{figure*}
{\centering {

\includegraphics[scale=0.4]{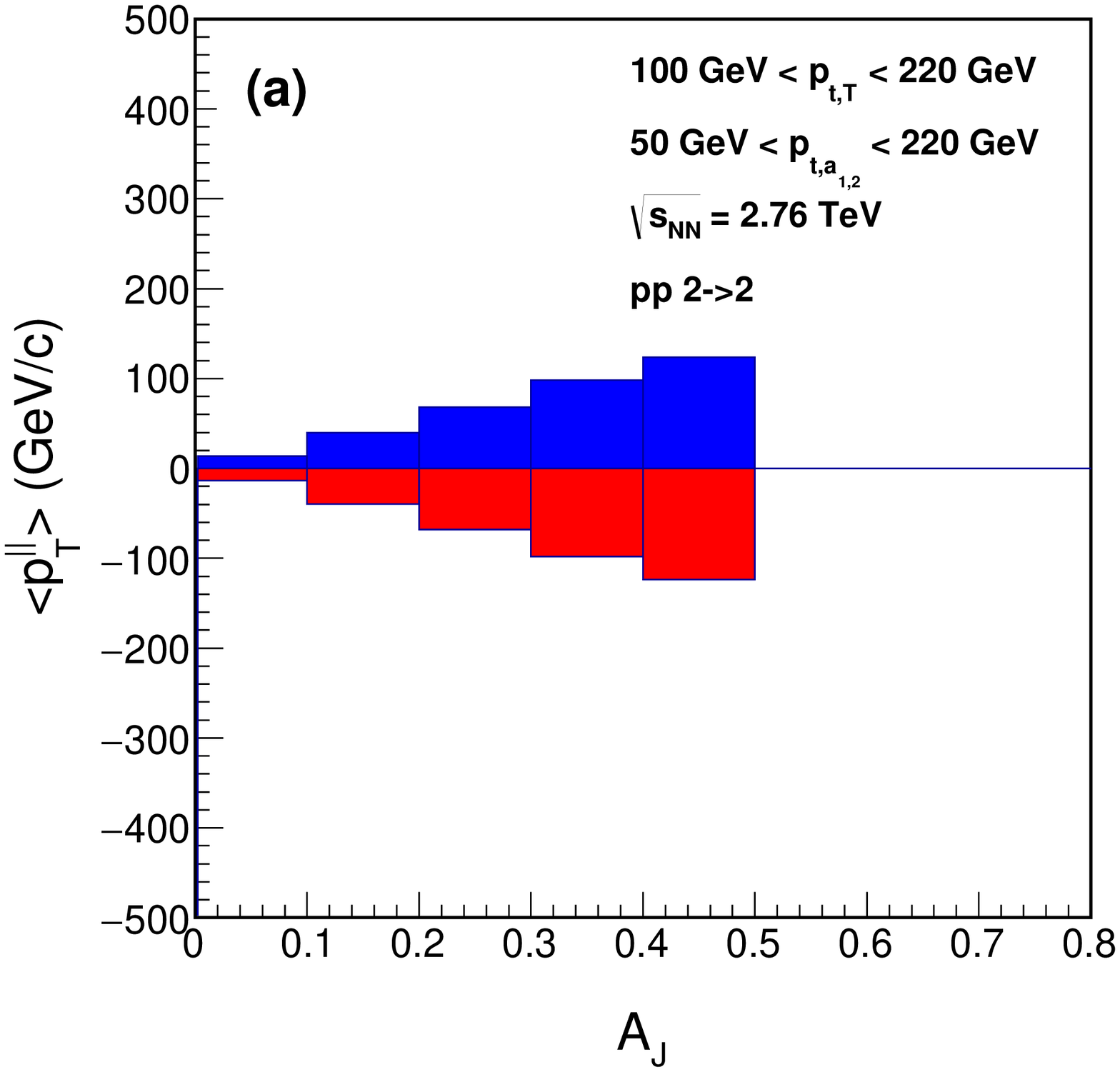} \includegraphics[scale=0.4]{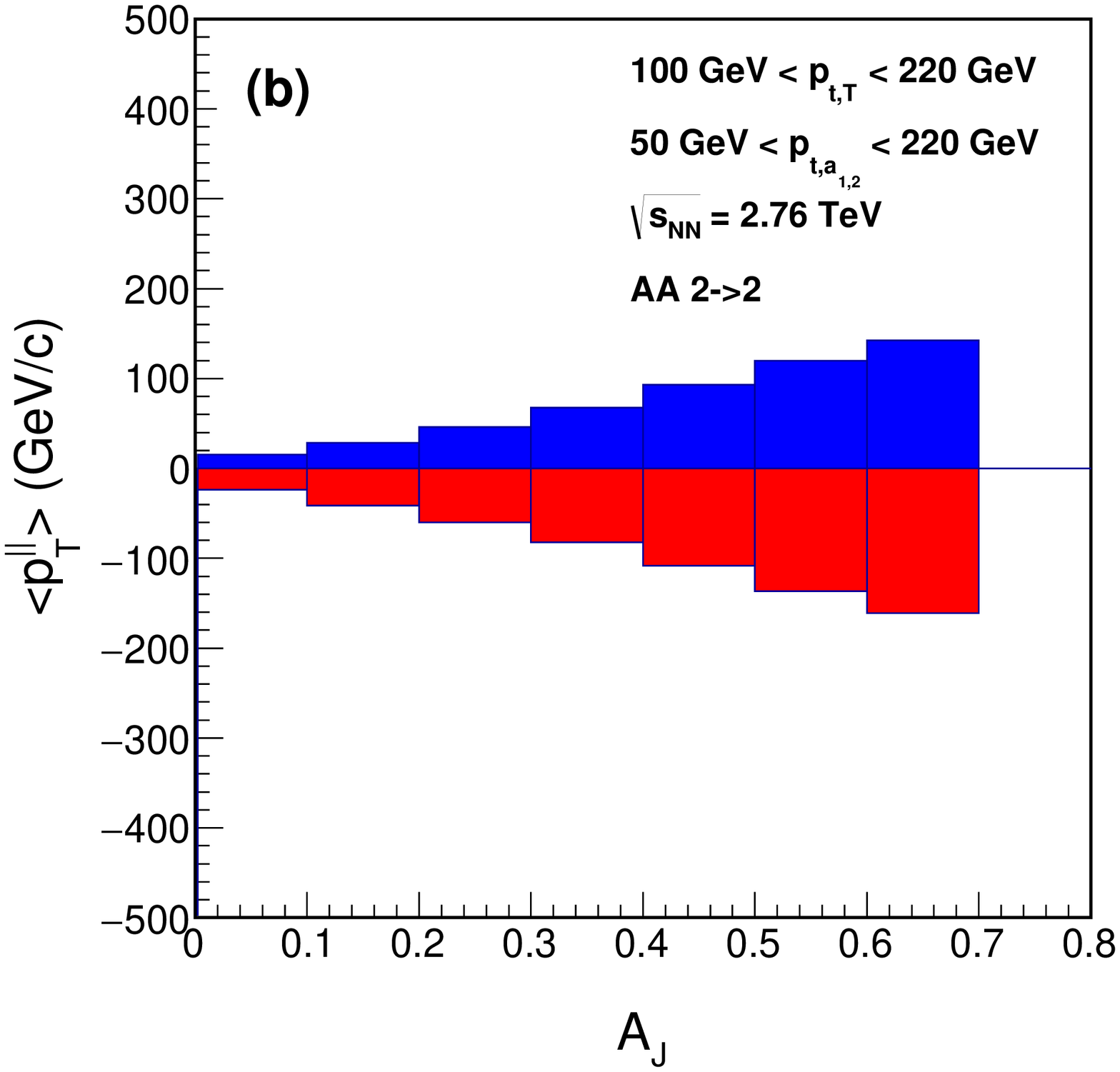}

\includegraphics[scale=0.4]{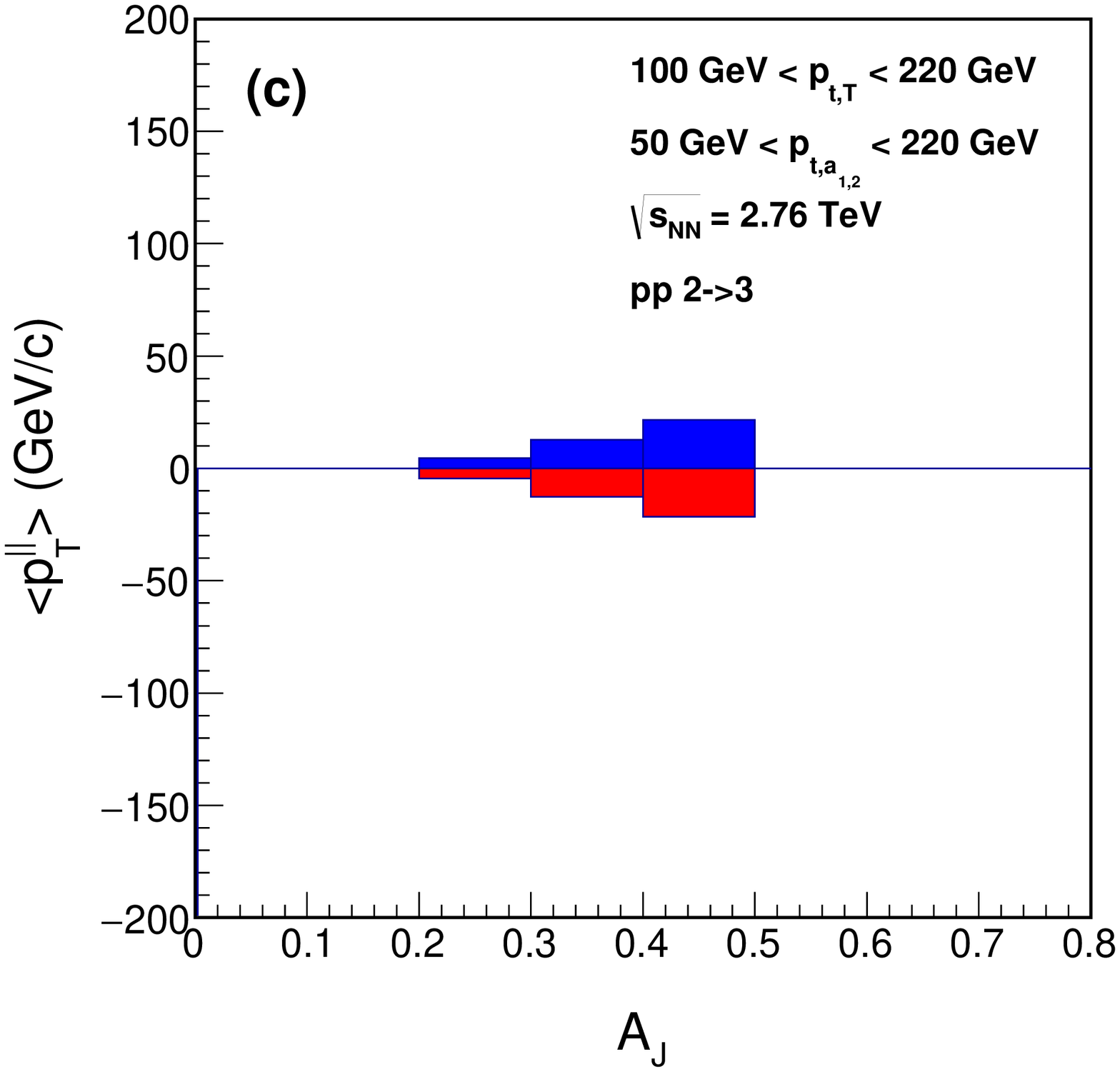} \includegraphics[scale=0.4]{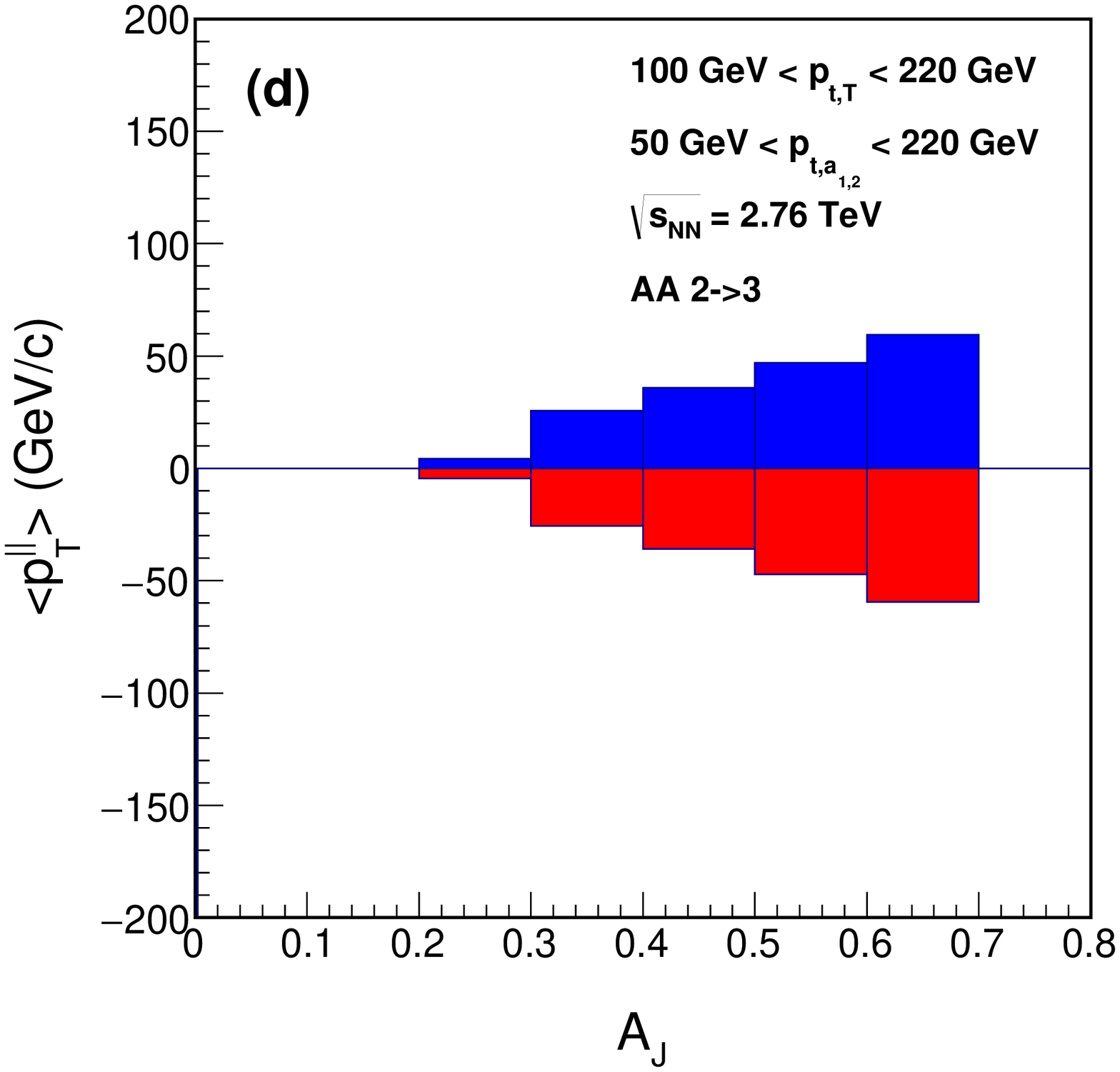}

}}
\caption{(Color online)
Average missing transverse momentum for leading jets of $p_{\mathrm T,1} > 100$ GeV and subleading jets of $p_{\mathrm T,2} > 50$ GeV, as a function of the event asymmetry. The left (right) column shows the balance for pp (AA) events arising from $2 \to 2$ (top row) and $2 \to 3$ (bottom row) partonic events. All panels show the negative momentum imbalance in the direction of the leading jet is shown above the horizontal axis and the momentum deposited into the medium is below the horizontal axis in the away-side hemisphere $\Delta\phi > \pi/2$. 
}
\label{fig3}
\end{figure*}

\begin{figure*}
{\centering {
\includegraphics[scale=0.4]{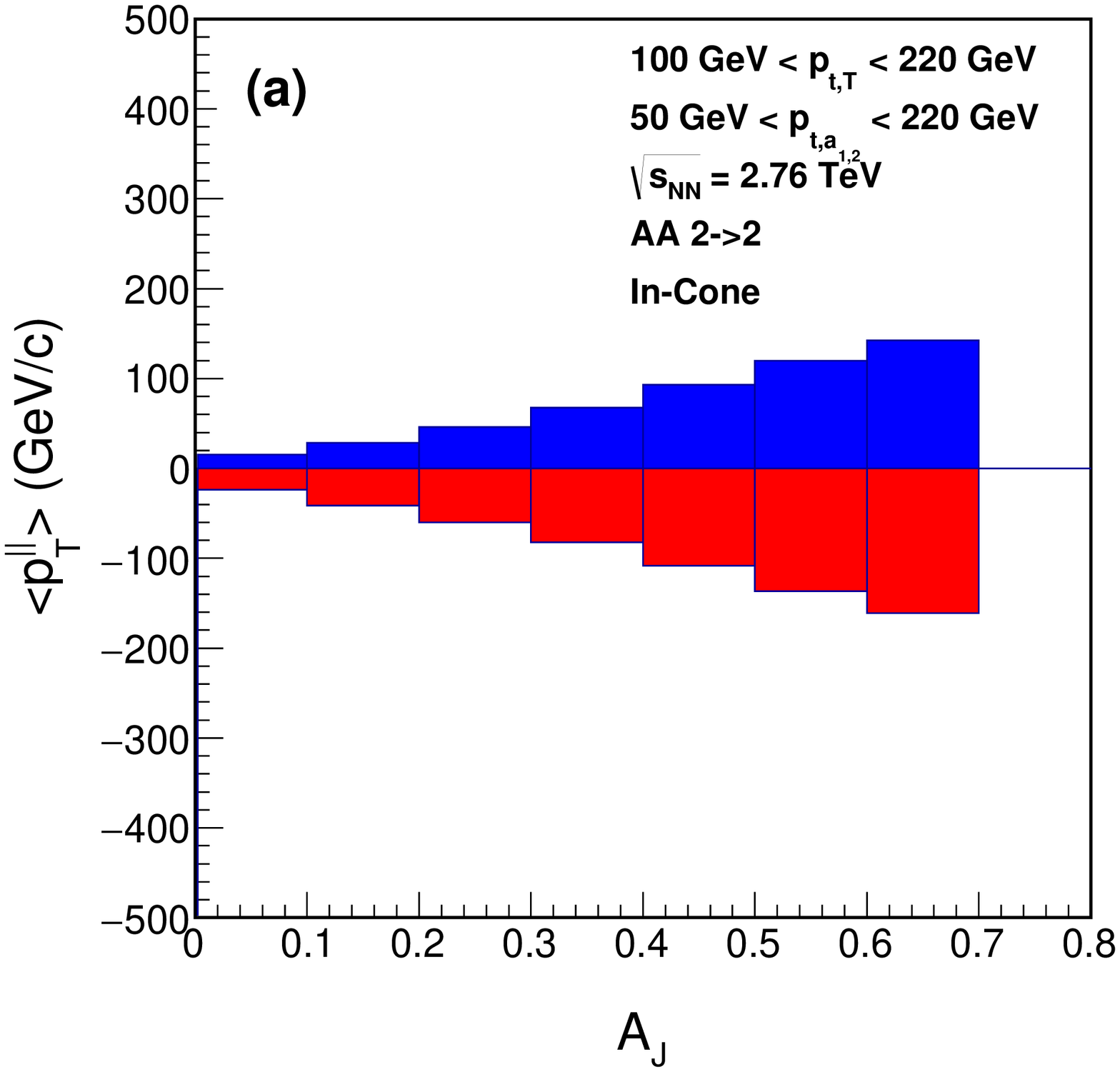} \includegraphics[scale=0.4]{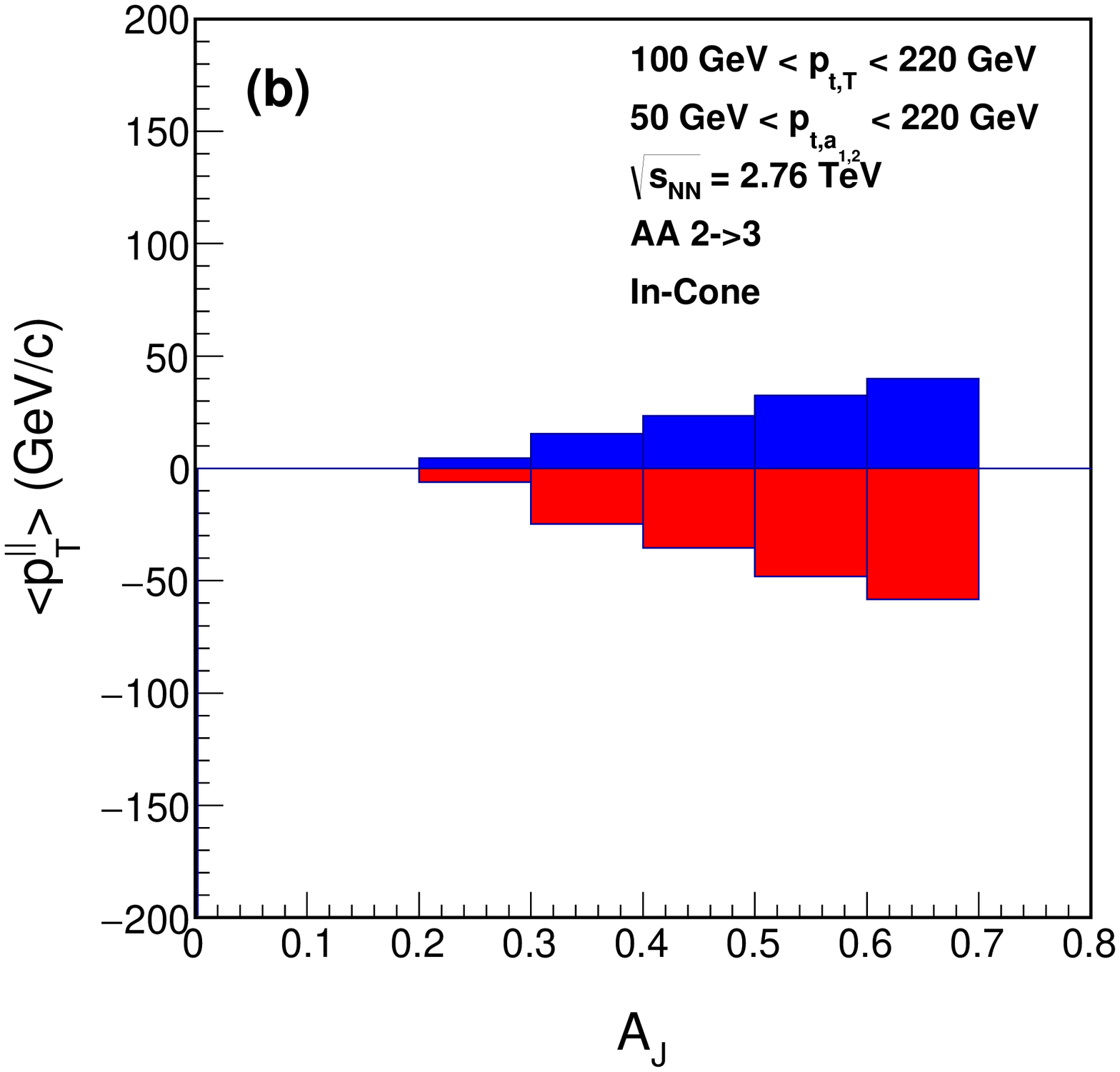}
}}
\caption{(Color online)
In-cone average missing transverse momenta for leading jets of $p_{\mathrm T,1} > 100$ GeV and subleading jets of $p_{\mathrm T,2} > 50$ GeV, as a function of asymmetry. The figure shows the in-cone balance for AA events arising from $2 \to 2$ (left panel) and $2 \to 3$ (right panel) partonic events. The negative momentum imbalance in the direction of the leading jet is shown above the horizontal axis and the momentum deposited into the medium is below the horizontal axis, inside a cone of $\Delta\phi > 3\pi/4$ ($R=0.5$).
}
\label{fig4}
\end{figure*}

\begin{figure*}
{\centering {
\includegraphics[scale=0.4]{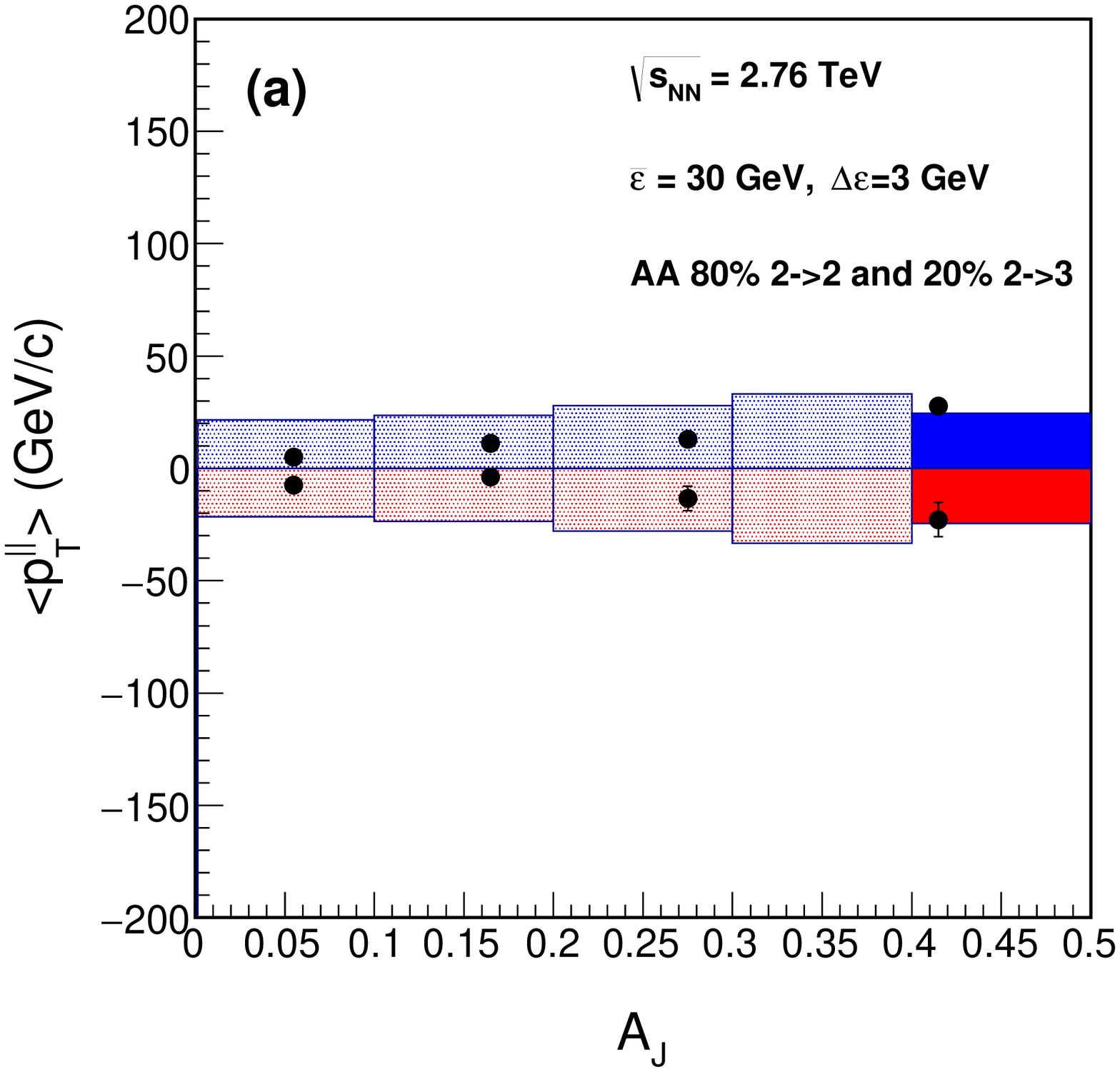}\includegraphics[scale=0.4]{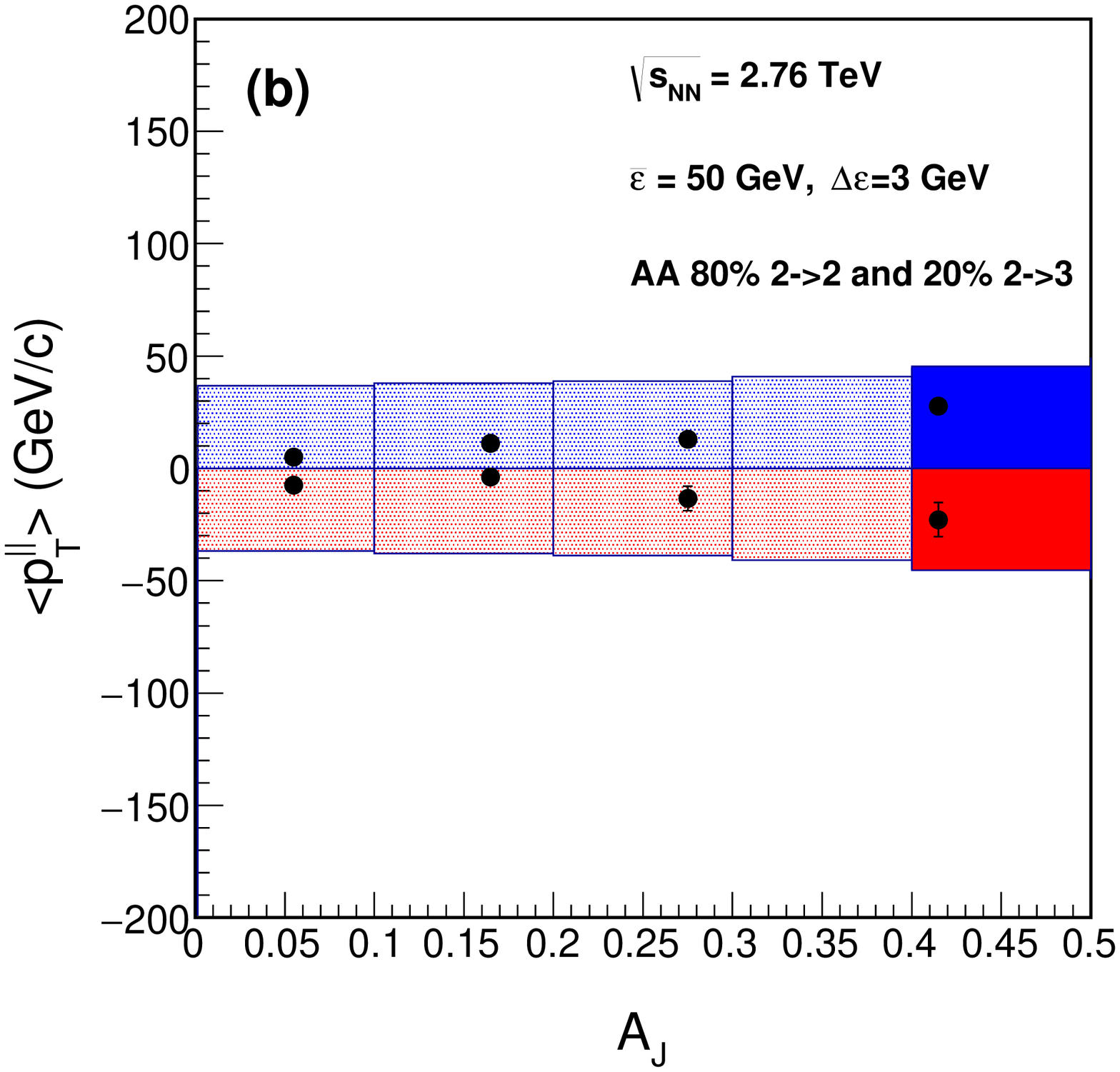}
}}
\caption{(Color online)
In-cone average missing transverse momenta for leading jets of $p_{\mathrm T,1} > 100$ GeV and subleading jets of $p_{\mathrm T,2} > 50$ GeV, as a function of asymmetry. The panels show the in-cone balance for AA events arising from $2 \to 2$ (shaded) with $\langle p_{\mathrm T}^{||} \rangle$ scaled to 80\% and the $2\to 3$ (solid) sample with $\langle p_{\mathrm T}^{||} \rangle$ scaled to 20\%. The sample used was obtained with the energy lost by a parton taken from a Gaussian distribution probability density $\rho({\mathcal{E}})$ with an average $\bar{\mathcal{E}}=30$ GeV 
and a half-width $\Delta {\mathcal{E}} = 3$ GeV (a) and with $\bar{\mathcal{E}}=50$ GeV and half-width $\Delta {\mathcal{E}} = 5$ GeV (b). The negative momentum imbalance in the direction of the leading jet is shown above the horizontal axis and the momentum deposited into the medium is below the horizontal axis, inside a cone of $\Delta\phi > 3\pi/4$ ($R=0.5$).
}
\label{fig5}
\end{figure*}

\section{Simulation}\label{II}

Our procedure consists of generating either $2\rightarrow 2$ or $2\rightarrow 3$ parton scattering events using Madgraph~\cite{madgraph}. In a given event we let one of the partons hadronize in vacuum~\cite{KKP} which is then labeled as the leading hadron, whereas the other partons travel through the medium and lose energy before they hadronize outside the medium.

Several energy loss scenarios have been considered to describe hadrons and/or jets emerging from a relativistic heavy ion collision. For example, in Refs.~\cite{renk,majumder} the authors make use of fragmentation functions, to help distinguish the stage of the collision where a given energy loss mechanism dominates. In this work we avoid implementing a specific mechanism but instead consider a simple and general scenario that encompasses many features of the ones discussed in the literature.

We consider a scenario where the hard scattering that gives rise to the final state partons happens near the medium's surface, such that the leading parton goes into vacuum whereas the loosing energy partons travel through most of the medium. To allow for an uncertainty in the location of the hard scattering we suppose that $\Delta E$ is normally distributed according to a probability density function.

We take the energy lost by a parton from a Gaussian probability density distribution $\rho({\mathcal{E}})$ with an average $\bar{\mathcal{E}}=50$ GeV and a width $\Delta {\mathcal{E}} =10$ GeV. This corresponds to a situation where the leading hadron is emitted near the surface. For definitiveness we consider that this happens at $\Delta L=5/6\ L$, where $L=10$ fm is the total medium's length and that the energy-loss per unit length as $dE/dx = 5$ GeV/fm. A parton with energy $E$ produced in a hard scattering will travel an average distance $\Delta L$ in the medium and lose energy $\Delta E$ (which on average is $\mathcal{E} \sim L (dE/dx) = 50$ GeV) to emerge from the medium with energy $E-\Delta E$ and fragment in vacuum. The above values emerge from a study of azimuthal particle distributions implemented in the context of energy deposition into the medium using linearized viscous hydrodynamics~\cite{Ayala3} that produces particles by means of a Cooper-Frye formalism. We assume that the sub-leading, but otherwise fast partons, produced in the hard scattering that travel through the medium do not change their direction of motion due to interactions with the medium. This seems to be a plausible assumption since the final hadrons, that these partons give rise to, are still very energetic. The interaction of the fast partons with the medium gives rise to energy and momentum deposited into the medium described by linear viscous hydrodynamics which in turn produces low momentum partons distributed around the fast ones according to the Cooper-Frye distribution.

Therefore, to compute $\delta p$, that is the average $p_{\mathrm{T}}$ deposited by the away-side parton into the medium, we write~\cite{Ayala3}
\begin{eqnarray}
  \left. \delta p \right|_{\phi^{\mbox{\tiny{max}}}} &=&\frac{\Delta\tau(\Delta y)^2}{(2\pi )^3}
   \int_{0}^{\infty}
   dp_T\frac{p_T^4}{T_0}\exp\left[ -p_T/T_0 \right]\nn\\
   &\times& 2 \int_{0}^{\phi^{\mbox{\tiny{max}}}} d\phi
   \left( \frac{\delta\epsilon}{4\epsilon_0} + \frac{{\mathbf{g}}_y\sin\phi + {\mathbf{g}}_z\cos\phi}{\epsilon_0(1+c_s^2)} \right).
   \label{intangledist}
\end{eqnarray}
The integral over $\phi$ in Eq.~(\ref{intangledist}) determines in- and out-of-jet cone regions where the average $p_{\mathrm T}$ is distributed and $\Delta y, \Delta\tau$ the rapidity and freeze-out time intervals, respectively.

The energy density $\delta\epsilon$ and the components for the momentum density ${\mathbf g}_i= \left(\bd{g}_T\right)_i + \left(\bd{g}_L\right)_i$ ($i=y,z$, $T$ and $L$ are the transverse and longitudinal modes) are given by
\begin{eqnarray}
  \delta\epsilon &=&
   \left(\frac{1}{4\pi}\right)
   \left( \frac{d E}{d x} \right)\left(\frac{2v}{3\Gamma_s}\right)^2 
   \left(\frac{9}{8v}\right)I_{\delta\epsilon}(\alpha ,\beta )
\label{moddeltaeps}
\end{eqnarray}
and
\begin{eqnarray}
\left(\bd{g}_{[T/L]}\right)_i &=&
   \left(\frac{1}{4\pi}\right)
   \left( \frac{d E}{d x} \right)\left(\frac{2v}{3\Gamma_s}\right)^2 I_{g_{[T/L]i}}(\alpha ,\beta ),
\label{modgtlyz} 
\end{eqnarray}
where the integrals $I_{g_{[T/L]i}}(\alpha ,\beta )$ are obtained from the solution of the linearized viscous hydrodynamic equations with a source term corresponding to the current produced by a localized and fast moving parton which is in turn proportional to a constant energy loss per unit length. These are given explicitly in Ref.~\cite{Ayala3} and written
in terms of the dimensionless variables
\begin{eqnarray}
   \alpha \equiv |z-vt|/\left(\frac{3\Gamma_s}{2v}\right) &~\mbox{and}~~& \beta \equiv x_T/\left(\frac{3\Gamma_s}{2v}\right),
\label{dimensionless}
\end{eqnarray}
which represent the distance to the source $|z-vt|$ and the distance
from the parton along the transverse direction $x_T$ (along the $\hat{y}$ axis
in the geometry we are using) in units of the sound attenuation length
\begin{eqnarray}
\Gamma_s\equiv \frac{4 \eta }{3\epsilon_0(1+c_s^2)}.
\label{defGammas}
\end{eqnarray}

As implemented in Ref.~\cite{Ayala3} $\epsilon_0$ is the static background's energy density and $c_s$ the sound velocity. This background is perturbed by the energy deposited by the fast parton moving near the speed of light. We focus on events at central rapidity $y\simeq 0$ and take the direction of
motion of the fast parton moving in the away-side to be the $\hat{z}$ axis (with the beam along the $\hat{x}$ axis). With this geometry, the transverse plane is the $\hat{y}-\hat{z}$ plane and therefore, the momentum four-vector
for a (massless) particle is explicitly given by $ p_\mu =
(E,p_x,p_y,p_z) = (p_T,0,p_T\sin\phi,p_T\cos\phi)$, where $\phi$ is the angle that the momentum vector ${\mathbf{p}}$
makes with the $\hat{z}$ axis.

Equation~(\ref{intangledist})  can be used in a two ways: on the one hand, by integrating the azimuthal angle over the full away-side hemisphere one can equate the average momentum deposited into the medium with the total energy lost by the fast parton and therefore extract the characteristic time $\Delta\tau$ for the process. In this way we get $\Delta\tau\simeq 8.5 ~\mbox{fm}$. On the other hand, given a certain amount of energy lost extracted randomly from the gaussian profile, by restricting the azimuthal angular integration, one can compute the amount of momentum going into the in- and out-of-cone regions around the away-side leading hadron.

The CMS collaboration supplemented their study with an analysis of the momentum imbalance inside and outside a cone with a fixed radius, as a function of the jet asymmetry $A_J$~\cite{CMS1, CMS2}. The analysis was done accounting for the different $p_{\mathrm T}$ contributions in each $A_J$ bin. The results showed that most of the momentum in the leading side is balanced with momenta in the 0 to 8 GeV/c range in the away-side, collected within a cone whose angular distance from the leading hadron was $\Delta\phi_{1,2} > 2\pi/3$. 

To account for the contribution of soft particles produced by the hadronization of the momentum deposited into the medium by the fast parton $\delta p$ in our model, we compute the \textit{in-medium} jet asymmetry variable as
\be
\label{asymm-in}
\tilde{A}_J= \frac{p_{\mathrm T,1} - \tilde{p}_{\mathrm T,2}}{p_{\mathrm T,1}+\tilde{p}_{\mathrm T,2}},
\ee
where $\tilde{p}_{\mathrm T,2} = p_{\mathrm T,2} + \delta p$,
when we are balancing the momenta inside the cone, with $\delta p$ given by Eq.~(\ref{intangledist}) and $\tilde{p}_{\mathrm T,2} = \Delta E - \delta p $, when we do it outside the cone. In the next section we look at observables plotted as functions of $A_J$ as defined in Eq.~(\ref{asymm}) for pp, or as functions of $\tilde{A}_J$ as defined in Eq.~(\ref{asymm-in}) for AA. Since what matters is the momentum of the final hadron and not the momentum of the initial parton used to compute the asymmetry, we simply call the asymmetry variable $A_J$.

\section{Results}\label{III}

In this section we present the results of the asymmetry and momentum imbalance analysis on the simulated samples, and compare this analysis to CMS data as reported in references~\cite{CMS1,CMS2}. We focus first on the asymmetry distributions for the $2 \to 2$ and $2 \to 3$ proton-proton and nucleus-nucleus samples together with a {\it combined} sample. The {\it combined} sample has 80\% of $2 \to 2$ prong events and 20\% of $2 \to 3$ prong events for both pp and AA.

Figure~\ref{fig1} shows the simulated dijet asymmetry distribution for leading jets of $p_{\mathrm T,1} > 100$ GeV and subleading jets of
$p_{\mathrm T,2} > 30$ GeV (left) and $p_{\mathrm T,2} > 50$ GeV (right).
The histograms show the benchmark asymmetry distribution for the pp $2\to 2$ sample (fine dotted line), the AA $2\to 2$ sample (dotted line) and the AA $2\to 3$ sample (solid line). Note that the pp $2 \to 2$ simulation is made out of symmetric events with $A_J$ in the range 0 to 0.3. On the other hand the AA $2 \to 2$ data shows a depletion of symmetric events and an enhancement of asymmetric ones. The shift in the peak and the widening of the distribution from pp to AA data, is purely due to the energy loss of the away-side parton that was implemented to generate this sample. The figure also shows a mutually complementary effect in the AA sample between the $2 \to 2$ and the $2 \to 3$ results: in the most symmetric bins ($A_j < 0.3$) the $2 \to 2$ events contribute more whereas in the most asymmetric ones ($A_j > 0.3$) the $2 \to 3$ events give a larger contribution as expected. The histograms on the left panel have a wider distribution than those on the right one which also have more symmetric events. This is caused by the lower cuts on $p_{\mathrm T}$ on the left side panels compared to more restrictive ones for the panels on the right.

Figure~\ref{fig2} shows simulated asymmetry distributions for leading jets of $p_{\mathrm T,1} > 100$ GeV and subleading jets of
$p_{\mathrm T,2} > 30$ GeV (left) and $p_{\mathrm T,2} > 50$ GeV (right). 
In the upper row we plot the asymmetry distribution for the benchmark pp $2\to 2$ sample and compare to CMS data for pp collisions in the 0-10\% centrality bin. The lower row shows the asymmetry distribution for the AA $2\to 2$ sample (solid line), the $2\to 3$ sample (fine dotted line) and the AA {\it combined} sample (dotted line) compared to CMS data for Pb-Pb collisions in the 0-10\% centrality bin~\cite{CMS1,CMS2}. There is a small difference in the description of the data when using the {\it combined} AA sample rather than the AA $2\to 2$, on the central asymmetry bins. The $2\to 3$ sample generates a slight depletion of symmetric events and a similarly small excess of asymmetric events. The interplay between this sample and the $2\to 2$ can account for some of the behaviour attributed to pure e-loss. The fact that we have a sample with 20\% of events in a $2\to 3$ configuration and are still able to describe the data supports the idea that a dijet sample with true $2 \to 3$ initiated dijet events can be a  component of the analysis that should be understood before coming to conclusions about the energy-loss mechanism at play.

The asymmetry distribution analysis can be complemented with a study of the average missing momentum as a function of jet asymmetry. This can help visualize the effect of the energy deposited into the medium for different cone radii. In our analysis it is crucial to see how much of the missing momentum for a given cone radius can be a attributed to the third hadron in a 2 to 3 event. Using Eq.~(\ref{ptmiss}) we can construct histograms showing the effect of energy loss both in the whole away-side hemisphere and only within a cone of size $\pi/2$ centered around the leading hadron, in the away-side.

Figure~\ref{fig3} shows the average missing transverse momentum for leading jets of $p_{\mathrm T,1} > 100$ GeV and subleading jets of $p_{\mathrm T,2} > 50$ GeV, as a function of the event asymmetry. The left (right) column shows the imbalance for pp (AA) events arising from $2 \to 2$ (top row) and $2 \to 3$ (bottom row) partonic events. All panels show the negative momentum imbalance in the direction of the leading jet (red histograms) and the momentum deposited into the medium (blue histograms) in the away-side hemisphere $\Delta\phi > \pi/2$. There are two main features worth noting in these plots: First, the pp sample fills up to $A_J = 0.5$ but the AA sample fills up to $A_J= 0.7$. Therefore there is a magnification on $\langle p_{\mathrm T}^{||} \rangle $ when going from pp to AA (both in the $2 \to 2$ and $2\to 3$ samples)  and this is due exclusively to the energy loss mechanism. Second, for the AA sample, the higher the $A_J$, the more important the $2\to 3$ component becomes with respect to the $2\to 2$ sample, with up to 30\% effect on the most asymmetric bins.

We now study the amount of missing $p_{\mathrm T}$ contained inside a cone of a given radius, as a complement of the half-hemisphere analysis done in Fig.~\ref{fig3}. Figure~\ref{fig4} shows the in-cone average missing transverse momentum for leading jets of $p_{\mathrm T,1} > 100$ GeV and subleading jets of $p_{\mathrm T,2} > 50$ GeV, as a function of asymmetry for AA events arising from $2 \to 2$ (a) and $2 \to 3$ (b) partonic events. Shown are the negative momentum imbalance in the direction of the leading jet and the momentum deposited into the medium inside a cone of $\Delta\phi > 3\pi/4$ ($R=0.5$). Note that the relative contribution of the in-cone momentum on the away side is smaller for the $2 \to 3$ as compared to the $2 \to 2$ case. This can be understood as a consequence of the second particle in the away side not being emitted within the reference cone. We emphasize that the analysis shown in Figs.~\ref{fig3} and~\ref{fig4}, agrees with the recent study of jet formation and bulk evolution interplay of Ref.~\cite{hirano}.

The CMS collaboration has reported an analysis whereby the balance of momentum is achieved by projecting all tracks onto the leading jet axis. From this sample a further selection of in- and out-of-cone tracks is made to study how much momentum is pushed outside the away-side cone, presumably by energy loss effects. We now make a comparison with CMS data and learn about the role the $2 \to 3$ processes have in the energy loss mechanism, which in turn will give us a rough estimate of their contribution to $\langle p_{\mathrm T}^{||} \rangle$. The comparison can only be done with balanced events at the hadronic level, so we implement a fragmentation where momentum is almost conserved (the fraction of hadron momentum from the parent parton is almost the same in the leading and the away sides), to emulate the fact that CMS has already projected all tracks onto the leading jet axis. Figure~\ref{fig5} shows the average missing transverse momentum for leading jets of $p_{\mathrm T,1} > 100$ GeV and subleading jets of $p_{\mathrm T,2} > 50$ GeV, as a function of asymmetry, arising from this AA sample, compared to CMS data~\cite{CMS1}. The histograms correspond to the $2\to 2$ (shaded) sample with $\langle p_{\mathrm T}^{||} \rangle$ scaled to 80\% and the $2\to 3$ (solid) sample with $\langle p_{\mathrm T}^{||} \rangle$ scaled to 20\%. The AA sample used was obtained with the energy lost by a parton taken from a Gaussian probability density distribution $\rho({\mathcal{E}})$ with an average $\bar{\mathcal{E}}=30$ GeV and a half-width $\Delta {\mathcal{E}} = 3$ GeV for panel (a) and with $\bar{\mathcal{E}}=50$ GeV and half-width $\Delta {\mathcal{E}} = 5$ GeV for panel (b). Note that a smaller average energy loss and a narrower width are more consistent with the data. Also, note that the most asymmetric bins in both panels are exclusively populated by the $2 \to 3$ sample, since the $2 \to 2$ sample cannot reach this asymmetry bin. This is due to the fact that with a balanced sample the asymmetry in the $2 \to 2$ case comes exclusively from energy loss. Therefore for the values  
$\bar{\mathcal{E}} \pm \Delta {\mathcal{E}}$ considered and the implemented $p_{\mathrm T}$ cuts, the largest asymmetry bins cannot be populated. This is not the case for the $2 \to 3$ sample, since one of the away side partons, could end up outside the reference cone.

\section{Summary and Conclusions}\label{Conclusions}

In this work, we have studied the way that fast moving partons deposit
their energy and momentum when travelling in a medium. We perform this study by looking into the momentum imbalance as a function of the jet asymmetry, using generated $2 \to 2$ and $2 \to 3$ parton events that lose energy and hadronize collinearly to form jets. To estimate the average energy loss deposited into the medium by the fast moving partons, we used linearized viscous hydrodynamics and the Cooper-Frye formula to calculate how this energy is distributed in different momentum contributions in the away-side. We argue that for conditions resembling those achieved in heavy-ion collisions, the shape of the obtained asymmetry and momentum imbalance agree with the ones reported by the CMS Collaboration. This in turn can be explained as originating from the energy loss of the partons created in the hard scatterings and travelling in the plasma, including a slight enhancement produced by the contribution from $2 \to 3$ events.

Furthermore, we found that the $\langle p_{\mathrm T}^{||} \rangle$ observable shows that the contribution of $2\to 3$ events is enhanced when going from proton-proton collisions to nucleus-nucleus collisions, with up to 30\% effect on the most asymmetric bins and agrees with the ones presented in recent studies of jet-bulk interplay \cite{hirano}. Finally, we perform an analysis with a balanced hadron momentum sample of the average missing $p_{\mathrm T}$ to compare with CMS data. We show that this analysis favours a smaller 
average energy loss and a narrower width, compared to the one used for our asymmetry distribution analysis. 

All together, our results suggest that an analysis containing a mixed two- and three-jet sample with a realistic energy-loss profile may prove useful for a better characterization of the jet momentum imbalance as a function of jet asymmetry. Progress in this direction will be reported elsewhere.

\section*{Acknowledgments}

Support for this work has been received in part from CONACyT-M\'exico under grant number 128534, from PAPIIT-UNAM under grant  number IN101515 and from {\it Programa de Intercambio UNAM-UNISON} and {\it Programa Anual de Cooperaci\'on Acad\'emica UAS-UNAM}. M. E. T.-Y. acknowledges support from the CONACyT-M\'exico sabbatical grant number 232946. J. J-M. is supported by the DOE Office of Nuclear Physics through Grant No.\ DE-FG02-09ER41620 and
by The City University of New York through the PSC-CUNY Research Award Program, grant 67732-0045.


\begin{thebibliography}{55}

\bibitem{CMS1} S. Chatrchyan {\it et al.}, (CMS Collaboration), Phys. Rev. C {\bf 84}, 024906 (2011).

\bibitem{CMS2} S. Chatrchyan {\it et al.}, (CMS Collaboration), {\it Measurement of momentum flow relative to the dijet system in PbPb and pp collisions at $\sqrt{s_{NN}}$ = 2.76 TeV}, CMS-HIN-14-010 (2014). 

\bibitem{renk} T. Renk, Phys.Rev. C {\bf 85}, 064908 (2012).

\bibitem{majumder} {\it Jet modification in the next decade: a pedestrian outlook}, A. Majumder, arXiv:1405.2019 [nucl-th].

\bibitem{others} J. Casalderrey-Solana, J. G. Milhano, U. A. Wiedemann, J. Phys. G {\bf 38}, 035006  (2011); G. -Y. Qin and B. Muller, Phys. Rev. Lett. {\bf 106}, 162302  (2011); Y. He, I. Vitev, B.-W. Zhang, Phys. Lett. B {\bf 713}, 224-232  (2012); C. Young, B. Schenke, S. Jeon, C. Gale, Phys.Rev. C {\bf 84}, 024907 (2011).


\bibitem{jetshydro} J. Casalderrey-Solana, E.V. Shuryak and D. Teaney, J. Phys.
Conf. Ser. 27, 22 (2005). Nucl. Phys. A 774 (2006) 57; A.K. Chaudhuri, U. Heinz, Phys. Rev. Lett. 97, 062301 (2006); R.B. Neufeld, B. Muller, J. Ruppert, Phys. Rev. C 78, 041901 (2008); B. Betz, M. Gyulassy, D.H. Rischke, H. Stocker, G. Torrieri, J.
Phys. G 35, 104106 (2008); R.B. Neufeld, T. Renk, Phys. Rev. C 82, 044903 (2010); B. Betz, J. Noronha, G. Torrieri, M. Gyulassy, D.H. Rischke, Phys.
Rev. Lett. 105, 222301 (2010).

\bibitem{jetsother} Y. Tachibana, T. Hirano; R.P.G. Andrade, J. Noronha, G.S. Denicol. arXiv:1403.1789 [nucl-th]; S. Floerchinger, Korinna C. Zapp, Eur.Phys.J. C74 12, 3189 (2014).

\bibitem{Blaizot} 
J.-P. Blaizot, Y. Mehtar-Tani, M. A. C. Torres, arXiv:1407.0326 [hep-ph].

\bibitem{hirano} Y. Tachibana, T. Hirano, Phys.Rev. C {\bf 90}, 021902 (2014).

\bibitem{Ayala1}
A. Ayala, J. Jalilian-Marian, J. Magnin, A. Ortiz, G. Paic, M. E. Tejeda-Yeomans, Phys. Rev. Lett. {\bf 104}, 042301 (2010); A. Ayala, J. Jalilian-Marian, A. Ortiz, Guy Paic, J. Magnin, M. E. Tejeda-Yeomans, Phys. Rev. C {\bf 84}, 024915 (2011).

\bibitem{Ayala2}
A. Ayala, I. Dominguez, J. Jalilian-Marian, J. Magnin and M. E. Tejeda-Yeomans Phys. Rev. C {\bf 86},  034901 (2012).

\bibitem{Ayala3}
A. Ayala, I. Dominguez, M. E. Tejeda-Yeomans, Phys. Rev. C {\bf 88}, 025203 (2013).

\bibitem{Ayala4} 
A. Ayala, J. D. Casta\~no-Yepes, I. Dominguez, M. E. Tejeda-Yeomans. arXiv:1412.5879 [hep-ph].

\bibitem{madgraph}
J. Alwall {\it et al.}, JHEP {\bf 1106}, 128 (2011).

\bibitem{KKP}
B.A. Kniehl, G. Kreimer and B. Potter, Nucl. Phys. B {\bf 582}, 514 (2000).

\end{thebibliography}
\end{document}